\documentstyle[11pt,newpasp,twoside,epsf]{article}
\def\edcomment#1{\iffalse\marginpar{\raggedright\sl#1\/}\else\relax\fi}
\reversemarginpar

\begin{document}
\title{The Secondary Stars in Short Period (Magnetic) Cataclysmic Variables}
 \author{Steve B. Howell}
\affil{University of California, Riverside, CA 92521}

\begin{abstract}
We review the recent knowledge base pertaining to causes for 
mass accretion variations (high/low states) in
highly magnetic cataclysmic variables  (i.e. polars). We then
examine what theory has to say about the properties of the 
secondary stars (the mass donors) with respect to those CVs with very short
orbital periods. Finally, we use recent observations of EF Eri in an extended low
state to provide direct observational evidence which shows
that this CV, as well as a few others, likely contain very low mass,
sub-stellar secondary stars.
\end{abstract}

\section{Introduction}
Mass accretion in cataclysmic variables has been extensively studied
in terms of their accretion disk and the theorized relation between
mass transfer rate and orbital period. Of course, the mass that is
transfered comes from the companion star and its rate, as well
as its character, are highly related to its evolutionary status.
In the past few years, studies aimed at direct observation of the
secondaries in CVs have been undertaken in an attempt to confirm theoretical
predictions for these stars..

For the short period CVs, the topic under consideration herein, there
are few cases of direct observation of their secondary, limited mostly
to brief glimpses during eclipse or serendipitous observation during
low states in magnetic systems. Purposeful observational programs
which use target of opportunity time to observe magnetic CVs during low
state have been started and they offer great hope in terms of providing
detailed spectroscopic views of the mass donors.

Low states, times of reduced mass transfer from the secondary star,
are an accepted observational property of CVs which contain white dwarfs
harboring a high magnetic field. The systems we speak of are called polars
or AM Herculis binaries and the white dwarf fields are 10-200 MG in strength.
It is probably the case that intermediate polars and non-magnetic CVs
have low states too, but they would be unobservable to the casual observer
due to the presence of the accretion disk (King \& Cannizzo 1998).

We briefly examine below, the present day ideas as to a cause for these
low states, actually for the entire suite of high, low, and in between
mass transfer states. We then look at what theory has to say about 
the mass donors in both the pre- and post-orbital period minimum systems. Finally,
we give some specific details about the polar EF Eri, lately sky-rocketed
into stardom due to its seven year low state and powerful observational
evidence for a brown dwarf-like secondary star.

\section{What is the Cause for High/Low State Behavior?}

Starting with initial works such as those of Bianchini (1992) and Hessman
et al. (2000),
evidence begin to emerge that seemed to imply a possible relation
between mass accretion variations (in particular low states) and solar cycle
type phenomena. The evidence showed $\sim$decade variations in overall
system brightness and theoretical support for star-spots migrating to the L1
region. Coupled with rapidly rotating, single late-type star
observations showing intense magnetic cycles, all indications were 
that CV secondaries should be extremely magnetically active and 
exhibit ''solar cycles".

Recently, low state observations of polars have shown fairly direct evidence
for star-spots on their surface (Howell et al. 2000) or TiO absorption
attributable to star-spots (Webb et al. 2002). Hessman (this volume)
has attempted to detail a possible mechanism
which may be responsible for the changing magnetic structures and
thus the observed mass transfer variations.

Based on the observational evidence presented in Howell et al., (2000)
for the polar ST LMi while in a low state, a theoretical model has emerged.
Noting the lack of ellipsoidal variations from the secondary star during
ST LMi's low state, Howell et al. calculated some basic parameters related
to the necessary conditions to fulfill mass transfer requirements. Using the
basic assumptions outlined in the standard model for CV evolution (see
Howell et al. 2001) and relatively simple physical arguments about what
is needed to provide the normal mass transfer rate  as a function of
orbital period, a toy model was produced.

\begin{figure}
\plotone{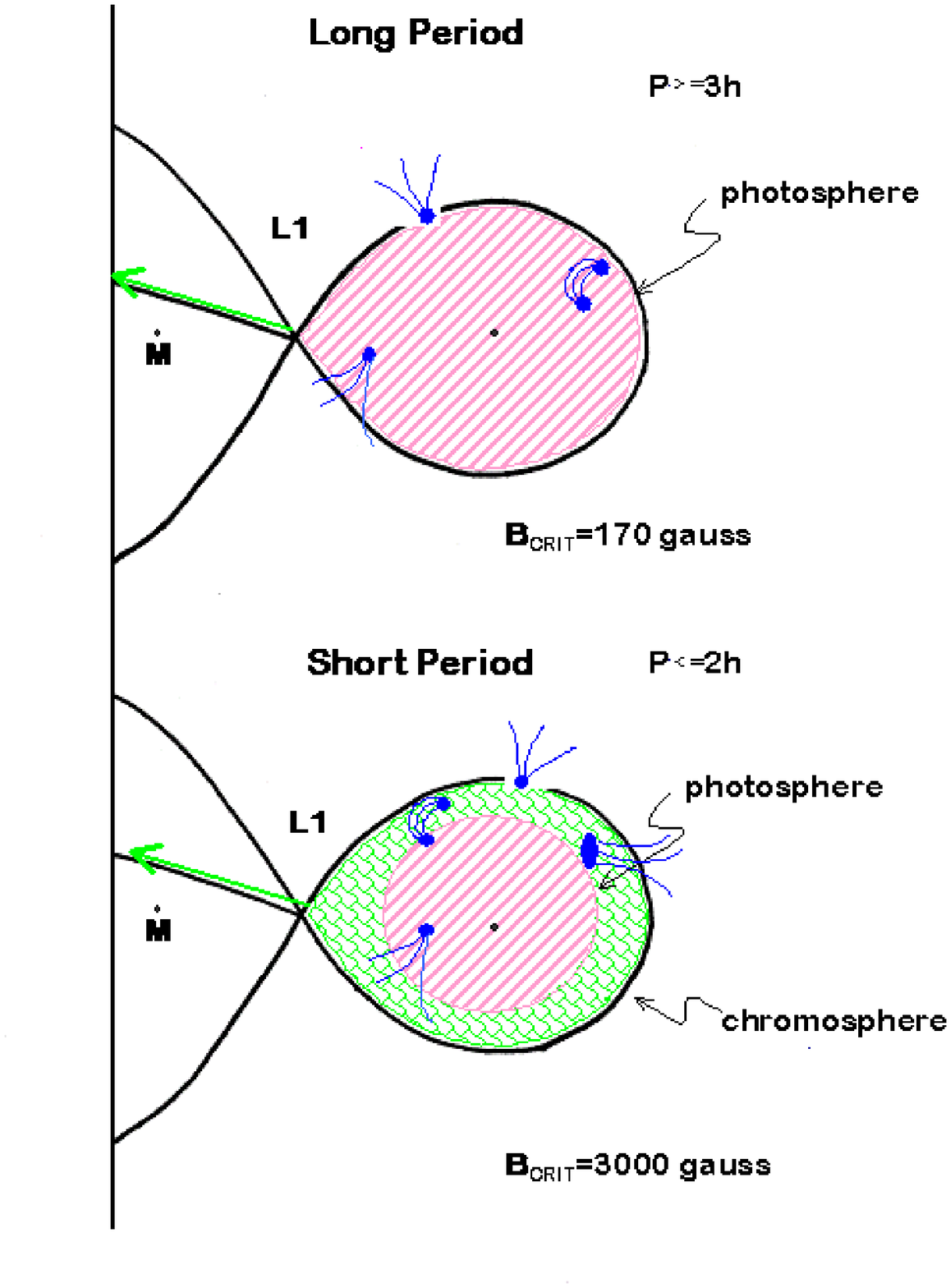}
\caption{A cartoon view of the toy model proposed by Howell et al., (2000)
for general mass transfer in CVs. Long period systems fill the Roche Lobes
with their photosphere while short period CVs need only their chromospheres
filling the Roche volume. See text for details.}
\end{figure}

Figure 1 provides the basic ideas formulated in this model. For long period
CVs (P$_{orb}\ge$3 hr), the photosphere must be in contact with the Roche
Lobe in order to provide the necessary mass flow. If the secondary star,
even locally, has a field strength of 170 gauss or more, magnetic lines
of force can and will remove material from the binary leading to
a stellar wind, in agreement with orbital angular momentum loss
due to magnetic braking. This field strength is rather modest and it
is probable that essentially all long period CVs meet this criteria.
For short period CVs, those with periods below the period gap,
the toy model makes very different predictions. First, the star must have
have a fairly substantial field strength ($>$ 3000 gauss) to
enable magnetic wind momentum loss. Secondly, the stellar photosphere
has such a high density, that its contact with the Roche surface
is not needed to provide the required mass transfer rate. For these
low mass stars (M$_{sec}<$0.25M$_{\odot}$), a standard chromosphere
provides the requisite density necessary to satisfy mass transfer needs.
The field strength of 3000 gauss or more may seem unlikely, but for such
a rapidly rotating star, extrapolation from single star values appears to
easily provide the needed field strength.

For a single late M star, stellar activity cycles are known to
provide chromospheres which can ``puff up" by 10 to 30 \% of the stellar
radius (Guinan 1999). Thus, a secondary star in a short period CV could easily
be imagined to fill its Roche Lobe with a chromosphere during a high state
(the toy model presented in Howell et al. assumes the chromosphere
to fill 10\% of the Roche Lobe volume during a high state)
and then to shrink back inside, thus decreasing or stopping mass transfer
during a low state. These modulations would therefore be highly correlated with
magnetic activity cycles in the star.

Recent observations of the pre-CV V471 Tau (Rottler et al. 2002)
also appear to provide evidence for stellar dynamo modulation of
magnetic cycles on the secondary star leading to changes in
the level of chromospheric activity.
While all the above ideas related to some form of
stellar activity mechanism similar
to solar cycles are appealing, the use of plane-parallel, non-rotating,
normal stellar atmosphere models backed by few observations are, at best,
only a starting point. Proper
interior and atmospheric models of the secondary stars in CVs and additional
phase resolved spectroscopy of the secondaries are needed.

\section{Theoretical Assumptions about the Secondary Stars}

Keeping to our belief that the ``standard model" for CV evolution
is valid, we find that for CVs with periods below the period gap,
it is possible for them to be pre- or post-orbital period minimum systems.
If pre-orbital period minimum, the secondary stars are essentially normal
stars in terms of their mass-radius relation but probably differ
from normal main sequence stars in terms of their atmospheric structure,
chemical composition, and effective temperature for a given mass.
Relations between their mass, radius, and orbital period are
presented in Howell et al. (2001).

For post-orbital period minimum CVs, the secondary stars are degenerate. 
That is,
they have masses less than $\sim$0.06 M$_{\odot}$, no core energy generation, 
and
stellar structures of a completely unknown kind. These secondary stars
are akin to brown dwarfs in some ways, but one must keep in mind that they
formed very differently from field brown dwarfs.

The relations between the mass, radius, and orbital period for post-orbital
period minimum CVs are:

\begin{equation}
M_{don} = 0.2053 - 0.2131 \cdot P_{orb} + 0.08762 \cdot P_{orb}^2
- 0.01282 \cdot P_{orb}^3
\end{equation}

\begin{equation}
R_{don} = 0.08195 + 0.01441 \cdot P_{orb} - 0.0002 \cdot P_{orb}^2
\end{equation}

\begin{equation}
R_{don} = 0.1439 - 1.703 \cdot M_{don} + 16.80 \cdot M_{don}^2
\end{equation}

\noindent
where mass and radius are in solar units and orbital period is in hours
(see Howell et al. (2001).

While no observational confirmation of these relationships (as well as
the orbital period - effective temperature relationship; see Mennickent
\& Diaz 2001) are as yet absolute, we will see below that the secondary
star in EF Eri and a few others seem to be strong candidates for post-orbital
period minimum CVs.

\begin{figure}
\plotfiddle{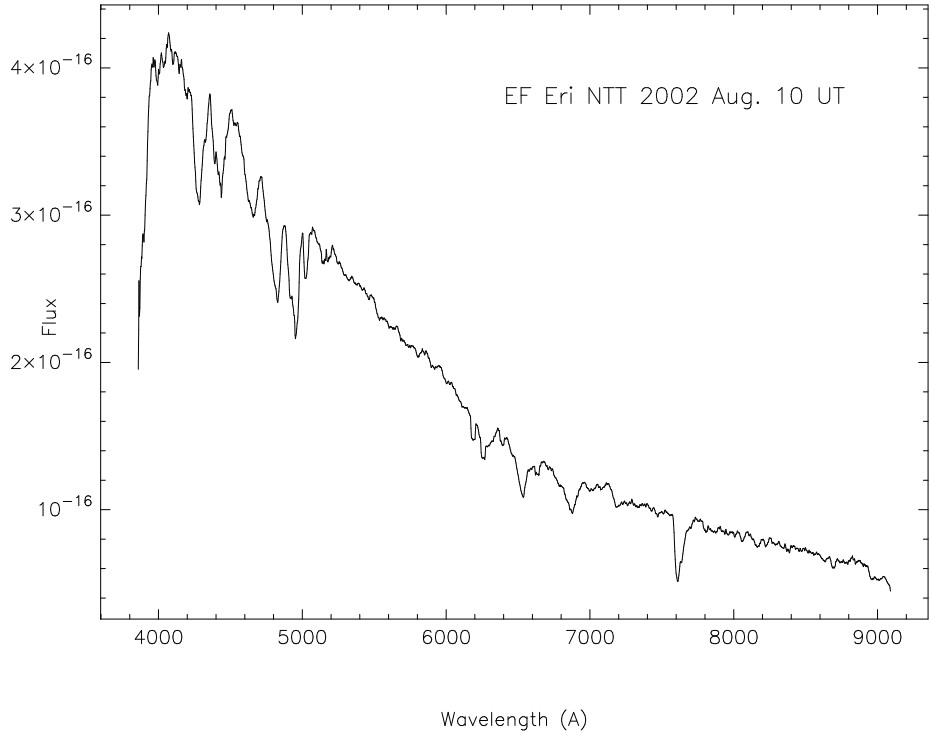}{3.6in}{-90}{55}{55}{-230}{320}
\caption{EF Eri observed at the NTT in August 2002. This
spectrum, seven years into its current low state, reveals no emission
lines or indication of residual cyclotron emission and clearly shows
the Zeeman split Balmer lines. Compare this spectrum with one of
similar wavelength coverage shown in Beuermann et al., (2000), 3 years into
the present low state, which has H$\alpha$ in emission and cyclotron
modulation in the near-IR. The spectral energy distribution in the NTT spectrum
is fit fairly well by a 9500K white dwarf.}
\end{figure}

\section{EF Eri}

EF Eridanus has been in a low state for over seven years now. Optical and IR
photometry and spectroscopy obtained over the past year provide an interesting
picture for the secondary star. The optical photometry shows a low level
modulation (0.1 mag) as the heated polar region gets self-eclipsed by the white
dwarf. Optical spectroscopy (Figure 2) shows nothing more than a cool, magnetic
white dwarf with all of the usual signs of mass transfer being absent.

\begin{figure}
\plotfiddle{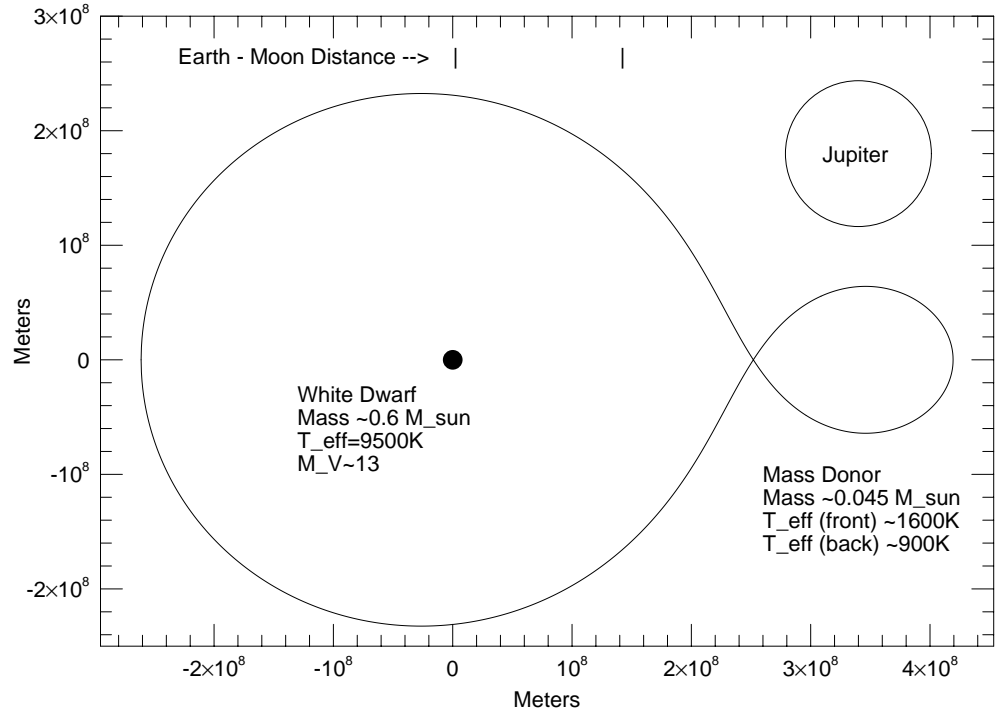}{3.5in}{0}{55}{55}{-230}{-45}
\caption{EF Eri shown to scale with the Harrison et al., (2003)
parameters listed. Additionally, the size of the planet Jupiter and
the Earth-Moon distance are drawn in. Note that even the white dwarf
size can be shown to scale in this very compact 81 min orbital
period binary.}
\end{figure}

Infrared photometry reveals in J, a similar light curve structure to that 
seen in the optical while H and K bands are not only completely different and
opposite in phase, but they show a single (note no ellipsoidal variations are
observed) large modulation of nearly 1 mag.
IR spectroscopy obtained at UKIRT and recently at Gemini North, show 
a secondary star which has an apparent temperature on the white dwarf facing
hemisphere of $\sim$1800K but is only about 900K on the 
opposite side. These temperatures
roughly correspond to an M8V star and a L8V brown dwarf.
Assuming M$_{don}$=0.045 M$_{\odot}$, EF Eri is
a cataclysmic variable containing a brown dwarf-like secondary. 

Detailed interpretation of these data and model results for EF Eri's component
stars are presented in Harrison et al. (2003). Figure 3 illustrates a
schematic model of the EF Eri system showing the particulars as well as 
its size in relation to the planet Jupiter and to the Earth-Moon distance.

\section{Summary}

Low state observations of the polar EF Eri in the optical and particularly 
in the infrared have revealed that it contains a very odd, low mass secondary
star. An object of seemingly unique characteristics with completely unknown
interior and atmospheric structure. More observations are needed of EF Eri
itself, as well as other candidate CVs which may contain similar odd mass 
donor stars.
Polars, during their low states, provide an ideal opportunity to 
directly observe
the secondary star with little or no interference from accretion flux.
This is in contrast to IR spectral observations of even the shortest period, 
low mass transfer rate disk systems which have, to date, essentially thwarted
all efforts to obtain essentially any (good quality) 
spectra of their secondary star.

The list of good suspects which probably contain brown dwarf-like secondaries
includes, in addition to EF Eri, 
LL And, AL Com, V592 Her, WZ Sge, and 1RXS J1050-1404. All of these systems,
except EF Eri, are
TOADs (Howell et al. 1995) and therefore the remaining members of this class 
of CV should be regarded as likely candidates as well. 

\acknowledgements
The author wishes to thank Tom Harrison and Elena Mason for observational
camaraderie and Kami Leanord, Elaine Owens, and Gil Esquerdo for
their help with Fig. 1 which once again showed why astronomers
``don't do windows".

\end{document}